# Distinction between the Preneoplastic and Neoplastic State of Murine Mammary Glands by spin-echo NMR

(NMR/relaxation times/diffusion coefficient of water protons)

C. F. HAZLEWOOD*†§, D. C. CHANG*¶, D. MEDINA‡, G. CLEVELAND¶, and
B. L. NICHOLS*†§

*Departments of Pediatrics, †Physiology, and ‡Anatomy, Baylor College of Medicine; §Texas Children's Hospital, and ¶Department of Physics, Rice University, Houston, Texas 77025

**ABSTRACT   We have, using spin-echo nuclear magnetic resonance spectroscopy, measured the relaxation times and diffusion coefficient of water protons in primary mammary adenocarcinomas of mice. In our biological model, three morphological stages were defined: (a) mammary gland tissue from pregnant mice, (b) preneoplastic nodules, and (c) neoplastic tissue. It was found that neoplastic tissues could be distinguished from normal and prenoeplastic tissue. Spin-spin and spin-lattice relaxation times and the diffusion coefficient of water protons are increased in the neoplastic tissue relative to mammary gland tissue from pregnant mice and preneoplastic nodule tissue. These results suggested that one can use a pulsed NMR method to detect and even predict breast cancer.**

Damadian reported recently that the spin-lattice ($T_1$) and spin-spin ($T_2$) relaxation times[1] of water protons in tumors were distinguishable from those in normal tissue or in benign tumors (1). These studies used tumor cell lines (Walker sarcoma and Novikoff hepatoma) that have been serially transplanted for many years, and, therefore, have had ample opportunity to undergo biochemical and morphological changes associated with tumor progression. In order to establish whether the increased motional state of tissue water demonstrated by nuclear magnetic resonance (NMR) measurements exists in primary tumors or whether these changes are a consequence of a tumor progression seen only in highly transplantable tumor lines, we chose to look at primary mammary adenocarcinomas of the mouse.

The mouse mammary tumor preparation has several advantages that are useful for investigation of discriminating changes in neoplastic compared to normal tissues. First, one can define at least three morphological states of mammary tissue—normal, preneoplastic nodules, and neoplastic tissue (2, 3). The hyperplastic alveolar nodule is considered preneoplastic in the sense that it gives rise to mammary tumors earlier and with greater frequency than normal mammary tissue. In addition, most murine mammary tumors, but not all, arise from the hyperplastic alveolar nodule (2, 3). Thus, the preneoplastic nodule is in a state intermediate between normal and neoplastic. Second, mammary adenocarcinomas are easily induced in high frequency by various methods. Finally, a great deal of information is known about mammary tumors; thus, these data can be correlated. Therefore, we have extended Damadian's observations to include (i) primary tumors, (ii) preneoplastic tissue, (iii) the measurement of another parameter (the diffusion coefficients), and (iv) another tissue type, i.e., mammary gland.

---

[1] Abbreviations: NMR, nuclear magnetic resonance; $T_1$, spin-lattice relaxation time; $T_2$, spin-spin relaxation time; $D$, diffusion coefficient of water.

## MATERIALS AND METHODS

Normal mammary tissues were taken from the inguinal (no. 4) mammary glands of 17- to 19-day old pregnant Balb/c and C3Hf mice. Preneoplastic nodule tissues were taken from Balb/c mice bearing nodule outgrowth lines $D_1$ and $D_2$ (4), and from C3Hf mice bearing a C3Hf nodule outgrowth line. The tumor-producing capabilities have been well defined under various experimental conditions. Outgrowth line $D_1$ produces only 2% mammary tumors after 360 days, whereas line $D_2$ produces 45% mammary tumors after 360 days (4). The nodule outgrowth lines are morphologically similar to normal mammary tissues found in mid-to-late pregnant mice. Mammary tumors were taken as primary tumors arising in these nodule outgrowth lines.

Measurements of $T_1$, $T_2$, and $D$ (diffusion coefficient) for cellular water protons were made by the "spin-echo" technique (5, 6). The excised mouse tissue was cut into small sections and packed into a nuclear magnetic resonance (NMR) tube so as to insure random orientation. All measurements were made at a resonance frequency of 30.3 MHz and 25°C.

$T_2$ and $D$ were determined by method "A" of Carr and Purcell (6) (i.e., a 90°-180°-180° pulse train). The echo amplitude was analyzed by the equation:

$$h_2 = h_1 \exp[-(t/T_2) - (1/12)\gamma^2 G^2 D t^3]$$

where $h_1$ and $h_2$ represent the amplitude of the first and second echo, respectively, $t$ is the observation time or time-separation of the echoes, $\gamma$ is the gyromagnetic ratio for the proton, and $G$ is the applied magnetic field gradient. A linear plot of ln $(h_2/h_1)$ against $G^2$ yields $T_2$ from the $y$-intercept, while $D$ can be calculated from the slope of the fitting line.

$T_1$ was determined in a similar manner by use of a 90°-90°-180° pulse train. The echo amplitude $h$ increases as $t'$, the 90°-90° separation, increases. A value for the maximum echo amplitude $h_0$ was obtained with a $t'$ of at least twenty times $T_1$. The growth of the echo follows the equation:

$$h = h_o[1 - \exp(-t'/T_1)]$$

$T_1$ is determined from a linear plot of ln $[1 - h/h_o]$ against $t'$.

## RESULTS AND DISCUSSION

The results of our measurements of $T_1$, $T_2$, and $D$ of water protons in normal mammary glands, preneoplastic nodules, and neoplastic tissue are summarized in Table 1. The relaxation times and the diffusion coefficient of water protons in these tissues are much smaller than in pure water. This agrees with the notion that the water molecules within living cells are strongly influenced by the water-ion-macromolecular interaction (7-13). Table 1 also shows that the three morphological stages (normal, preneoplastic, and neoplastic) of mammary tissue can be distinguished by the average values of the physical parameters $T_1$, $T_2$, and $D$. Furthermore, it seems that there is a progressive change in these parameters as the mammary tissue advances toward the neoplastic state. The relaxation times of water protons in tumors are almost doubled relative to mammary gland tissue from pregnant animals and to nodule tissue. The relaxation times for the nodule (preneoplastic) tissue also appear to be slightly (but significantly) longer than those for mammary gland tissue from pregnant animals[2].

**TABLE 1**. *Relaxation times and diffusion coefficient of water protons in pure water, normal mammary gland, preneoplastic nodule outgrowth, and neoplastic tissues at 25°C* *

| Tissue and number of samples | $T_1$ (s) | $T_2$ (s) | $D$ (cm$^2$/s $\times$ 10$^{-5}$) |
|---|---|---|---|
| Pure water | 3.1 | 1.43 ± 0.27 | 2.38 ± 0.016 |
| Tumor (5) | 0.920 ± 0.047 | 0.091 ± 0.008 | 0.78 ± 0.05 |
| Nodule (5) | 0.451 ± 0.021 | 0.053 ± 0.001 | 0.44 ± 0.03 |
| Normal pregnant mammary gland (5) | 0.380 ± 0.041 | 0.039 ± 0.002 | 0.34 ± 0.04 |

* (*a*) The pure water data were from four samples of water except for $T_1$, where only two samples were analyzed. (*b*) All values for tissue-water protons are given as the average ± the standard error. (*c*) No significant differences were found in the three values ($T_1$, $T_2$, and $D$) measured within the nodule lines and within the tumors, the results were combined for each tissue group.

The diffusion coefficient shows the same trend as the relaxation times. That is, $D$ is increased in neoplastic mammary gland relative to nodule or normal mammary gland. Again, the diffusion coefficient for water protons in nodule tissues is slightly larger than that for normal mammary gland.

The lipid content is high in the mammary gland tissue from pregnant animals as well as in the nodule tissue. Most of the protons in this lipid are mobile enough so that they are detected by the NMR spectrometer. We found that the protons in the lipid fraction produce a high resolution NMR signal that is chemically shifted 205 Hz from the water proton signal. The presence of the lipid proton signal complicates the analysis of our pulse NMR data. First, it increases the uncertainty in the $T_1$ and $T_2$ measurements for nodule and mammary gland tissue from pregnant animals. Second, since water cannot penetrate the lipid layers, lipids can serve as discrete physical barriers and compartmentalize the water molecules. The measured diffusion coefficient of water protons, therefore, might be reduced because of the compartmentalization by the lipids (14), rather than by a specific water-protein interaction.

---

[2] The difference between the average values of relaxation times in nodule tissues and normal mammary gland gave a *P* value between 0.15 and 0.10. This difference might be more significant if we increased the number of samples analyzed.

In a pilot study, we have found the lipid content of mammary glands from 17-day old pregnant animals and of Balb/c nodule line $D_2$ to be about 23 and 17%, respectively. The lipid content of neoplastic tissue, on the other hand, was found to be much lower (1-3%). This reduction of lipid content may be one of the causes for the increase in the measured diffusion coefficient. However, it is doubtful that the difference in relaxation times can be explained by the changes in lipid content, since the exchange of lipid protons and water protons is negligible. The increase in the relaxation times for neoplastic tissue represents a change in the interaction of water molecules with macromolecular structures of the cell. A similar change in water interaction has also been observed in skeletal muscles of developing rats. The relaxation times decrease as the muscle tissue mature (15).

The above results, along with those of Damadian, support the hypothesis concerning water-macromolecule interaction in the oncogentic process advanced by Ling (ref. 12, chap. 18, p. 483) and, Szent-Gyorgyi (16).

We would like to acknowledge the following for their support: Robert A. Welch Foundation, The David Underwood Trust Fund, National Science Foundation, and USPH (Research Grants: RR5425, FR-00259, CA-11944, and RR-0188 from the General Clinical Research Centers Program of the Division of Research Resources, National Institutes of Health). We thank Dr. H. E. Rorschach (Chairman, Department of Physics, Rice University) for his constant encouragement, cooperation, and helpful discussions. The assistance of Mrs. Betty E. Perronne and Mrs. Shirley Lincoln is also acknowledged.